# Global Climate network evolves with North Atlantic Oscillation phases: Coupling to Southern Pacific Ocean

O. Guez[1(A)], A. Gozolchiani[2], Y. Berezin[1], Y. Wang[1] and S. Havlin[1].

[1] Department of Physics, Bar-Ilan University, Ramat-Gan 52900, Israel.
[2] Institute of Earth Sciences, The Hebrew University of Jerusalem, Jerusalem, 91904, Israel.

**Abstract** – We construct a network from climate records of atmospheric temperature at surface level, at different geographical sites in the globe, using reanalysis data from years 1948-2010. We find that the network correlates with the North Atlantic Oscillation (NAO), both locally in the north Atlantic, and through coupling to the southern Pacific Ocean. The existence of tele-connection links between those areas and their stability over time allows us to suggest a possible physical explanation for this phenomenon.

**Introduction.** – In recent years, complex network framework has been applied to analyze climate fields such as temperature and geopotential height at a certain pressure level. The aim of this is to follow climate dynamics and to investigate the temporal stability of their structure. The climate network approach has been applied to study the temporal evolution and global imprints of El-Nino, North Atlantic Oscillation (NAO) and Rossby waves [1–14].

An empirical orthogonal function analysis reveals that the North Atlantic Oscillation is the most dominant mode of variability of the surface atmospheric circulation in the Atlantic [15]. The oscillation is present throughout the year in monthly mean data, but its most pronounced stage occurs during the winter [16]. Both phases of the NAO are associated with basin-wide changes in the intensity and location of the North Atlantic jet stream and storm track [17], and in large-scale modulations of the normal patterns of zonal and meridional heat and moisture transport, which in turn results in changes in temperature and precipitation patterns often extending from eastern North America to western and central Europe [18,19].

In this paper we present a new application of the network approach to locate areas around the globe that are influenced by the NAO. We begin by describing the method for construction of the climate network, including the data and the numerical procedure. Next by analyzing the dynamics of the climate network we show that NAO events might interact with very far locations from the north Atlantic basin, in the South Pacific Ocean. Finally we summarize and discuss the possible implications of our findings, and connect them to a known physical mechanism relating the two basins.

**Methods.** –

*Data.* We analyze the National Center for Environmental Prediction/National Center for Atmospheric Research (NCEP/NCAR) reanalysis air temperature fields at 1000hPa [20] and the ERA-40 reanalysis [21] . For each node of the network, daily values for the period 1948-2010 are used, from which we extract anomaly values (actual values minus the climatological averaged over the years for each day).

(A)E-mail: oded.guez@biu.ac.il

The data is arranged on a latitude-longitude grid with a resolution of $2.5° \times 2.5°$.

In our work we analyze a regional network and a global network. The regional network is located in the regime which is most relevant for NAO (shown in fig. 2) [22]. The pressure difference between the Icelandic-Low and the Azores-High pressure centers, which are included in our regional network, is frequently used as the NAO Index [23]. In the regional network there are 33 nodes in the east-west direction and 25 nodes in the north-south direction, amounting to total of 825 nodes. For the global network (shown in fig. 8) we choose 726 nodes covering the globe in an approximately homogeneous manner.

*Numerical procedure*. Similarly to earlier studies [3,11,12], we define the weight (strength) of the link measured from a date $y$ on, connecting the nodes *m* and n as:

$$W_{m,n}^y = \frac{MAXC_{m,n}^y - MEANC_{m,n}^y}{STDC_{m,n}^y}. \tag{1}$$

Where MEAN is the average, STD is the standard deviation and MAX is the maximal value of the absolute value of the cross covariance function $C_{m,n}^y(\tau)$[1]. We further define the time lag, $\tau_{m,n}^y$, as the shift (in days) corresponding to the highest peak of $C_{m,n}^y(\tau)$. We chose to analyze only the winter season (December of the current year to March of the following year) when the NAO effects are known to be stronger [24]. We measure time lags in the range between -72 and +72 days. This interval is chosen to be long enough so that $W_{m,n}^y$ is not sensitive to our choice.

We define the total weighted degree of a node as the sum of the weights of all links which attach to this node:

$$T_m^y = \sum_n W_{m,n}^y, m \neq n \tag{2}$$

---

[1] This cross covariance function is computed from data that begins in the year represented by y and ends in y+2. This choice is made in order to have sufficient statistics as well as to capture the dynamical changes of these correlations. Upper limits of $y$ such as $y + 1, y + 3$ yield similar results but less pronounced.

**Results.** – Here we explore a relation between the structure of the climate network and the NAO events.

*The relation between NAO and the total weighted degree of nodes in the regional network*. Fig. 1 shows the total weighted degree of nodes in each year, $T_m^y$. One can observe a region of dominant nodes (Index 400-650) that correlate with a running average[2] NAO-Index: the total weighted degree (upper panel) obtains high values when the NAO-Index (lower panel) is positive and vice versa. This behavior suggests a positive correlation [25] between the number of strong links in the network and the NAO-index, in agreement with ref. [12].

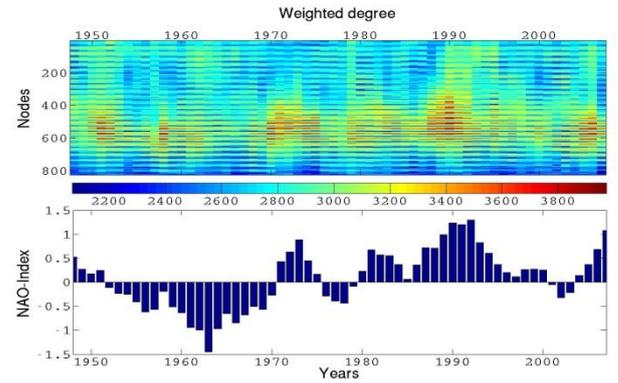

Fig. 1: (Colour on-line) Upper panel: the total weighted degree of nodes in each year, $T_m^y$, in the regional network (described in the methods section). The colours (see bar code) represent the strength of the weighted degree nodes. Lower panel: The smooth NAO-Index, $I^y$.

In order to locate the geographical location of the highly connected nodes, seen in fig. 1, we first calculate the mean of the total weighted degree of nodes over the years, $T_m$ (fig. 2). One can see that highly weighted degree nodes are localized in the east side of the Atlantic Ocean, near the east coast of US. In order to identify whether a link reflects a physical coupling rather than a random noise, and in order to avoid boundary affects on our result, we create a surrogate network [26]. To achieve this, we shuffle the data such that it preserves all the statistical quantities of the data, as the distribution of values, and their autocorrelation properties, but omits the physical dependence between different nodes. The network properties in such a case are only due to the statistical quantities and therefore are similar in their properties to spurious links in the original network. To follow this strategy, we choose for each node a random sequence of years where the order of the days within the year is preserved.

---

[2] We smooth the data using a moving average filter (bandwidth = 5 months). Moving average is a lowpass filter with filter coefficients equal to the reciprocal of the span. We do so since our climate is also an average as described in [1].

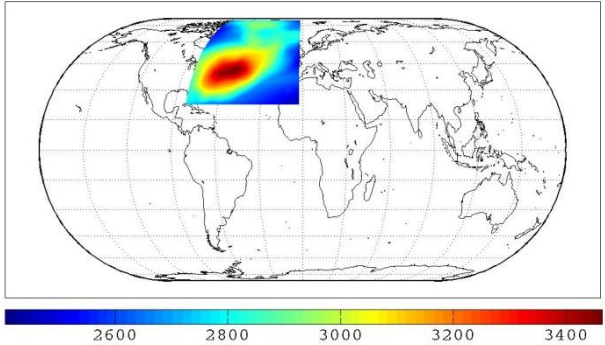

Fig. 2: (Colour on-line) The mean, $M_m$, over the years of the total weighted degree of nodes, $T_m^y$, in the regional network. The colours (see bar code) represent the strength of the weighted degree nodes.

We show in fig. 3 the histogram of the mean over all the years of the total weighted degree of nodes, $T_m^y$, for real and shuffled data. It is evident that the distribution of weighted degrees of the surrogate data has a cut-off around $T_m = 2,650$. Node degrees above this level are therefore not likely to emerge due to random fluctuations.

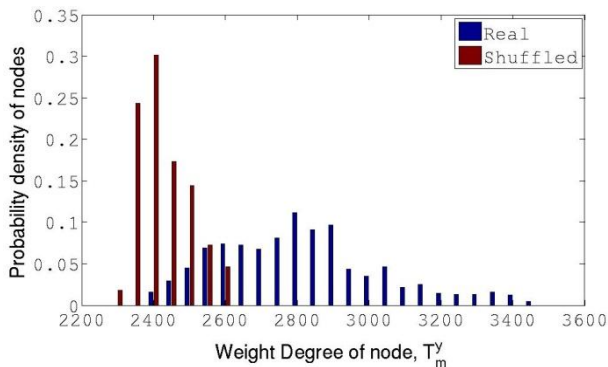

Fig. 3: (Colour on-line) The histogram of the mean total weighted degree of nodes, $T_m$, in the regional network. Comparison to surrogate shuffled data is also shown (The statistic two-sample Kolmogorov-Smirnov test is 0.73 and asymptotic p-value is 0).

Next we calculate the correlation, $C_m$, between the total weighted degree of nodes, $T_m^y$, and of the NAO-Index, $I^y$, and show this in fig. 4. One can see that nodes with high $C_m$ values are located between latitudes ($45° - 60°N$ in the center of the Atlantic Ocean.

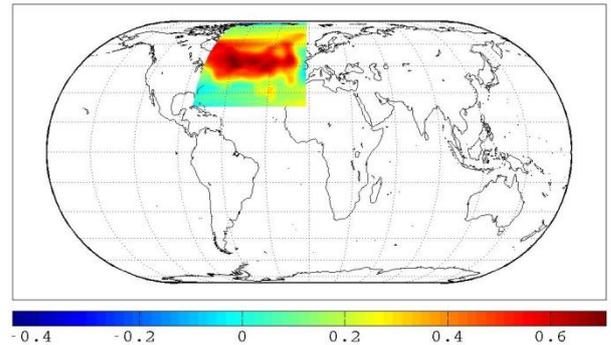

Fig. 4: (Colour on-line) The Correlations, $C_m$, between the total weighted degree of nodes, $T_m^y$, and the NAO-Index, $I^y$, in the regional network. The colours (see bar code) represent the correlation of the weighted degree nodes with NAO.

In fig. 5 we show the distribution of $C_m$ from our surrogate data, along with the real data. The regime $C_m > 0.45$ is evidently above noise level.

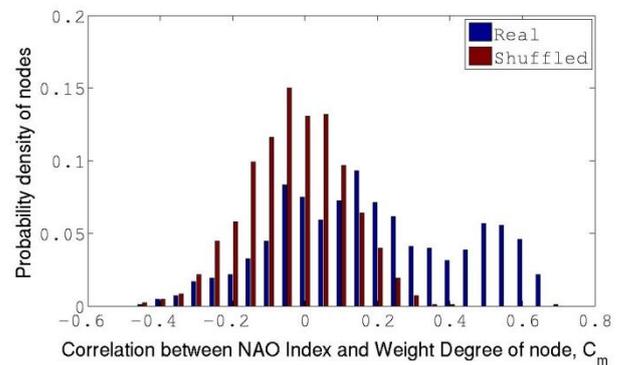

Fig. 5: (Colour on-line) The histogram of correlations, $C_m$, between the total weighted degree of nodes, $T_m^y$, in the regional network, and the NAO-Index, $I^y$. Comparison to surrogate shuffled data is also shown (The statistic two-sample Kolmogorov-Smirnov test is 0.43 and asymptotic p-value is 0).

We summarized the results of our two types of shuffled tests in table 1. It is evident from the table and from fig. 3 that the majority of our calculated weighted degree values, $T_m^y$, are above noise level (633 nodes, 77% of total 825 nodes). On the other hand, the statistical significance test for correlation values, $C_m$, is much more strict and only 22% of our nodes are found to be significantly correlated with the NAO index. To conclude, the significance test for correlation values, $C_m$, is strict enough to also insure that the weighted degree values, $T_m^y$, are above significance level.

(A)E-mail: oded.guez@biu.ac.il

O. Guez *et al.*

Table 1: The distribution of nodes with regard to the significance thresholds of weighted degree and correlation.

| Number of nodes (% of total 825 nodes) | $C_m < 0.45$ | $C_m > 0.45$ |
|---|---|---|
| $T_m < 2,650$ | 184 (22%) | 8 (1%) |
| $T_m > 2,650$ | 459 (56%) | 174 (21%) |

*The interaction between NAO and the total weighted degree of nodes in the global network.* Next we quantify the interactions of the NAO zone with other zones around the globe. To this end, we repeat the analysis of previous sections using a sparse global grid of 726 nodes. To avoid the influence of random links shown in this network, we pick a suitable threshold for link weights. In fig. 6 we plot the histogram of link weights in the regional network, $W_{m,n}^y$. One can estimate from the histogram of the surrogate data that $W_{m,n}^y > 5[StD]$ is above noise level.

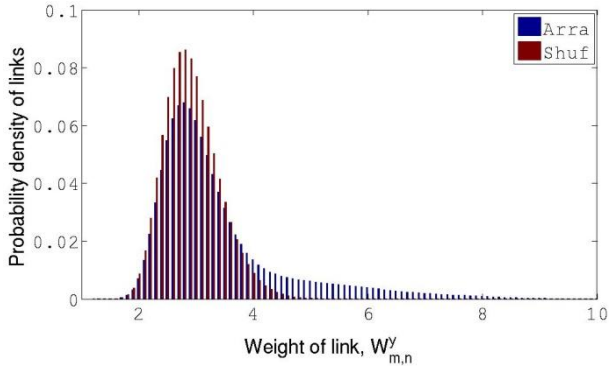

Fig. 6: (Colour on-line) The histogram of the weight of pairs over all the years, in the regional network (The statistic two-sample Kolmogorov-Smirnov test is 0.17 and asymptotic p-value is 0).

Filtering out links which are below significance level we calculate the correlation, $C_m$, between the total weighted degree of nodes, $T_m^y$, and the NAO-Index, $I^y$, for the global network and show this in fig. 7. As found earlier, in this paper, the dynamics of node degrees in the North Atlantic is highly correlated with the NAO index. However, further higher (in absolute value) anti-correlations exist between NAO and the southern Pacific Ocean. To investigate this phenomenon, we show those nodes in the southern Pacific Ocean in fig. 7 and refer them as anticorrelated nodes *(ACN)*.

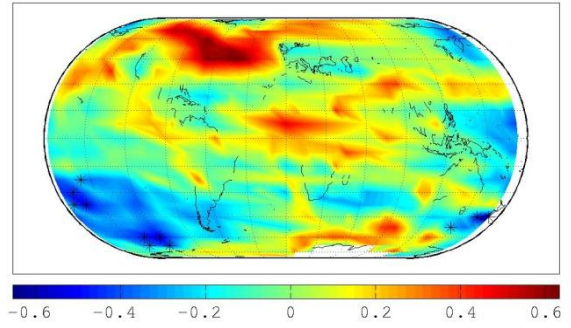

Fig. 7: (Colour on-line) The map of correlations, $C_m$, between the total weighted degree of nodes, $T_m^y$, in the global network and the NAO-Index, $I^y$. In black stars (*) is The locations of nodes in the southern Pacific Ocean which acquire high correlations (in absolute value).

Each one of these ACN nodes has both short range connections with neighboring nodes and long ranged tele-connections with distant nodes in the northern hemsiphere, and we show those long ranged tele-connections in fig. 8 for two typical nodes. Most of them are stable and exist during all years.

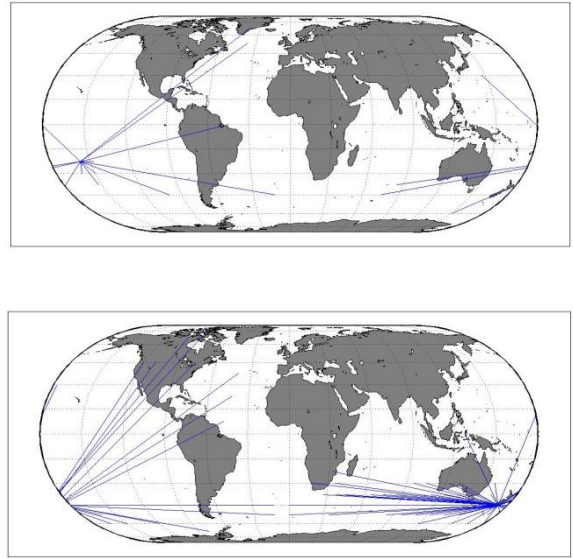

Fig. 8: (Colour on-line) The links of typical node in the southern Pacific Ocean for a selected year 1948-1951. The upper panel is the node in location $(22.5°S, 202.5°W)$ and the lower panel is the node in location $(45°S, 172.5°E)$.

Since we investigate the coupling of North Atlantic Oscillation and Southern Pacific Ocean, thus we choose to focus on the stable (i.e. the link exist more than 20 years of the 59 years possible) and long range (distnce of link larger than 5,000 Km) links, and caculating the mean value of the time-delay during those years. We show in fig. 9 that those links represent a very fast signal, crossing between hemispheres in typicaly 1 to 4 days and the mean value is up to 3 days. There are more links with negative time-delay (the link is from shoutern node to northern node) than links with negative time-delay (the link is from northern node to shoutern node). One can observe mostly two length scales collections of links: one around 7,000 Km and the other around 14,000 Km, which indicate of a relation to the one and two wavelengths of the observed Rossby wave (respectively) [14].

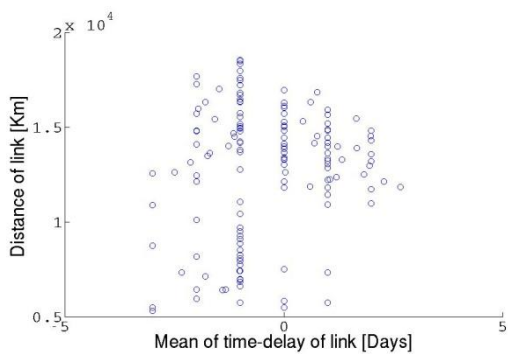

Fig. 9: The scatter plot of the distances between links versus the mean value of time-delay of stable links (exist more than 20 years of the 59 years possible) for the ACN nodes in the southern Pacific Ocean.

Finaly, a possible physical explanation for this pehenomena is suggsted as follows: The Antarctic Oscillation (AAO) is the dominant pattern of non-seasonal tropospheric circulation variations south of 20S, and it is characterized by pressure anomalies of one sign centered in the Antarctic and anomalies of the opposite sign centered about 40-50S [27]. The variability of the Antarctic Oscillation Index (AAO) is associated with anomalous zonal wind in the upper troposphere over the central-eastern tropical Pacific. Rossby waves, high altitude fast waves moving to the west, often follow an irregular path compatible with the region of AAO activity. Anomalous northward flux of energy across the tropics due to the quasi-stationary Rossby waves, is known to be trigger the Pacific/North American (PNA) and the North Atlantic (NA) teleconnection patterns in the hnorthern hemisphere. This mechanism shows up in the northern hemisphere in the form of dynamical pressure patterns. The time scale of the delay of links can vary from -5 to +5 days, as seen in fig. 9.

**Summary.** – We have constructed climate networks based on surface air temperatures data in the North Atlantic region, and in the entire globe. We find that variability of weighted degrees in the north Atlantic closely follows the NAO index. On the global map we observe a large southern Pacific region with negative correlation between its weighted degree variability and the NAO index. Evaluating these nodes reveal a robust pattern of links crossing between the hemispheres in 1-4 days. These links are in line with a known anomaly wind pattern described in [27].

∗ ∗ ∗

The authors would like to acknowledge the support of the LINC project (no. 289447) funded by the EC's Marie-Curie ITN program (FP7-PEOPLE-2011-ITN) and the Israel Science Foundation.

(A)E-mail: oded.guez@biu.ac.il